\documentclass[aps,twocolumn,showpacs,amsmath,amssymb,floatfix]{revtex4}
\usepackage{graphicx}
\usepackage{dcolumn}
\usepackage{bm}
\usepackage{longtable}

\begin{document}

\title{Radiation forces on an absorbing micrometer-sized sphere in an
evanescent field}
\author{I. Brevik}
\email{iver.h.brevik@mtf.ntnu.no}
\author{T. A. Sivertsen}
\email{Tom-Arne.Sivertsen@ffi.no}
\affiliation{Department of Energy and Process Engineering,\\Norwegian University of
Science and Technology,N-7491 Trondheim, Norway}

\author{E. Almaas}
\email{Almaas.1@nd.edu}
\affiliation{Department of Physics, \\University of Notre Dame, Notre Dame, Indiana
46556}

\begin{abstract}
The vertical radiation force on an absorbing micrometer-sized
dielectric sphere situated in an evanescent field is calculated, using
electromagnetic wave theory. The present work is a continuation of an
earlier paper [E. Almaas and I. Brevik, J. Opt. Soc. Am. B {\bf 12},
2429 (1995)], in which both the horizontal and the vertical radiation
forces were calculated with the constraint that the sphere was
non-absorbing. Whereas the horizontal force can be well accounted for
within this constraint, there is no possibility to describe the {\it
repulsiveness} of the vertical force, so distinctly demonstrated in
the Kawata-Sugiura experiment [Opt. Lett. {\bf 17}, 772 (1992)],
unless a departure from the theory of pure non-dispersive dielectrics
is made in some way. Introduction of absorption, {\it i.e.} a complex
refractive index, is one natural way of generalizing the previous
theory. We work out general expressions for the vertical force for
this case and illustrate the calculations by numerical
computations. It turns out that, when applied to the Kawata-Sugiura
case, the repulsive radiation force caused by absorption is {\it not}
strong enough to account for the actual lifting of the polystyrene
latex or glass spheres. The physical reason for the experimental
outcome is, in this case, most probably the presence of surfactants
making the surface of the spheres partially conducting.
\end{abstract}

\maketitle

\section{Introduction}

The outcome of the Kawata-Sugiura levitation experiment from 1992
\cite{kawata92} is in some ways still surprising. The authors examined
the radiation force on a micrometer-sized spherical dielectric
particle in the evanescent field produced by a laser beam of moderate
power, $P=150$ mW. As one would expect, the particle moved along the
horizontal plate at a speed of a few micrometers per second, as a
result of the tunneled photons in the evanescent field. The surprising
point is, however: why does the particle {\it lift} from the surface?
As reported in Ref.~\cite{kawata92}, ``the particle is forced to float
from the substrate surface and to slide along the surface.'' For a
non-magnetic and non-absorbing dielectric particle, it is well known
that the electromagnetic volume force density ${\bf f}$ is
\begin{equation}
{\bf f}=-\frac{1}{2}E^2 {\bf \nabla} \epsilon, \label{1}
\end{equation}
$\epsilon$ being the permittivity such that ${\bf f}=0$ in the
homogeneous and isotropic interior, and different from zero only in
the particle's boundary layer. The evanescent field is known to vary
as $\exp[-\beta (x+h)]$ in the vertical direction (see Eq.\ (\ref{10})
below), and is significant only very close to the substrate. In
practice, this field interacts with those parts of the particle that
are situated closest to the substrate, and it gives rise to a surface
force that, according to Eq.\ (\ref{1}), has to act in the {\it
downward} direction (assuming that $\epsilon > \epsilon_0$). It is
simply impossible to account for a lift force on the particle, using
Eq.\ (\ref{1}). This discrepancy between electromagnetic theory for
dielectrics and the Kawata-Sugiura experiment \cite{kawata92} is
therefore, as mentioned above, somewhat surprising at first sight,
since Eq.\ (\ref{1}) has generally been proven to be invaluable in a
multitude of cases in electromagnetism and optics.

Consequently, we have to conclude that the explanation for the
observed lift of the particle has to lie in the presence of an
additional force, different from the one given by Eq.\ (\ref{1}). The
following two possibilities come to mind:

(1) The particle has a complex permittivity $\epsilon$, and is
therefore exposed to a repulsive (or positive lift) force caused by
the absorption of radiation in the interior.

(2) Another natural possibility is that {\it surface effects} may have
played a significant role in the experiments. The presence of
impurities on the surface, or the presence of a liquid of a film of
adsorbed material, may have a considerable influence on the electrical
conductivity of the surface region, and can thus enhance multiple
scattering effects.  This allows for an analysis of the problem in
terms of a multiple scattering formalism. Adopting such a picture, the
vertical force on the particle necessarily has to be repulsive; the
photons are bouncing off the spherical surface and are, accordingly,
transferring an impulse in the upward direction.

In the present paper, we will be studying option (1) above, thus
allowing for a complex $\epsilon$, which in turn implies a
non-vanishing effective conductivity $\sigma$ in the interior of the
particle. As far as we know, such a study has not previously been
undertaken. This study is a continuation of the work of Almaas and
Brevik \cite{almaas95}, dealing with the case of a pure dielectric
(real $\epsilon$). Whereas the analysis in \cite{almaas95} gave
satisfactory results for the {\it horizontal} radiation force observed
in the Kawata-Sugiura experiment, it was impossible, as mentioned
above, on the basis of Eq.\ (\ref{1}) to account for an upward
directed vertical force. From now on, we will assume that the particle
is a sphere of radius $a$, and regard the effective conductivity
$\sigma$ as an input parameter.

Before embarking on the mathematical formalism, let us however make
some further remarks related to option (2) above.  There is a variety
of surface effects that one can envisage: multiple reflections, as
mentioned; moreover capillary forces (the order of magnitude of the
capillary forces may be up to $10^{-7}$ N), surface tension (because
of the capillary forces), viscous retarding forces if the sphere and
the surface are immersed in a liquid, electrostatic forces, and
finally the role of evaporation. Cf. in this context the survey given
by Vilfan {\it et al.} \cite{vilfan98}. Especially the role of the
{\it size} of the spherical particle seems to be important.  If the
particle is large, multiple scattering between the sphere and the
substrate distorts the evanescent wave structure that one would have
in the absence of the particle.  Lester and Nieto-Vesperinas
\cite{lester99} recently analyzed the problem using multiple
scattering methods, and were able to account for a positive lift
force. An advantage of this kind of method is that the proximity of
the sphere-substrate two-body system is addressed explicitly. In
contrast, our formalism is built on the model of scattering of an
evanescent wave by an isolated sphere, and is therefore not expected
to be very accurate for small separations between the sphere and the
substrate. It is necessary to point out that the adoption of a
multiple scattering method does not {\em in itself} imply that the
lift force is positive, as long as the material in the sphere is a
pure dielectric. Regardless of whether the electromagnetic boundary
conditions at the surface of the substrate are taken into account or
not, for a pure dielectric, the force on the sphere is given by Eq.\
(\ref{1}), and, as we have argued above, is necessarily downward
directed. To account for a positive lift force, we must allow for
deviations from the simple theory of pure dielectrics, meaning that
absorption, and consequently conductivity, are important effects in
the theory, explicitly or implicitly.

As already mentioned, we will not attempt to give a quantitative
description of a possible adsorbed film on the surface; the physical
reason for such a film may be quite complex.  Problems of this sort
belong to the vast field of physicochemical hydrodynamics. [A good
source of information in this area is the book of Levich
\cite{levich62}; a briefer treatment is found in Landau and Lifshitz
\cite{landau87}]. It is noticeable that the experiment of Vilfan {\it
et al.}  \cite{vilfan98} similarly measured a repulsive force. They
obtained a force 300 pN on a dielectric sphere of radius 5 $\mu$m in
the evanescent field of a 500 mW laser focused to a diameter of 5
$\mu$m. These spheres are relatively large, and are therefore likely to
distort the evanescent field structure considerably.

In the next section, we summarize the wave theoretical formalism
describing the interaction between the evanescent field and the
sphere. In Section 3, we show how the surface force ${\bf
F}^{\mathrm{surf}}$ -- {\it i.e.} the part of the total radiation
force ${\bf F}$ which is associated with the boundary and independent
of the absorptive properties in the interior -- can be written as a
double sum over complex expansion coefficients. Section 4, the main
section of our paper, contains the calculation of the absorptive part
${\bf F}^{\mathrm{abs}}$ of the radiation force. The results are
illustrated in Figs. \ref{fig:2} - \ref{fig:5}. Our calculations
incorporate both the two polarization states of the incident beam ($p$
and $s$ polarization).

Our main analytical result is given in Eq.\ (\ref{56}), for the
non-dimensional vertical absorptive force component
$Q_x^{\mathrm{abs}}$. As one would expect physically, the result
becomes proportional to the effective conductivity $\sigma$. For the
polystyrene latex spheres, or glass spheres, used in the
Kawata-Sugiura experiment, the values of $\sigma$ are low. Hence, it
turns out that $Q_x^{\mathrm{abs}}$ becomes too weak to account for
the floating of the sphere in the gravitational field. Now, our
formalism has, of course, a much wider scope than the Kawata-Sugiura
experiment. In optics, there are numerous cases where radiation forces
interact with absorbing media. For instance, the current development
of microelectromagnetical systems, which are used in a variety of
applications including even measurements of the Casimir force,
accentuates the usefulness of the present theory.

\section{Basic formalism}

\begin{figure}[t]
\includegraphics[width=5cm]{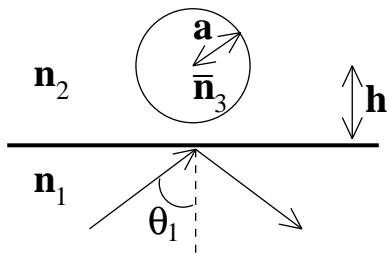}
\caption{Spherical particle of radius $a$, with complex refractive
         index $\bar{n}_3$, situated in an evanescent field with its
         center at a height $h$ above the plane substrate. A laser
         beam is incident from below at an angle of incidence
         $\theta_1 > \theta_{crit}$. Refractive indices in the
         transparent media 1 and 2 are respectively $n_1$ and $n_2$.}
\label{fig:1}
\end{figure}

A sketch of the geometry is shown in Fig. \ref{fig:1}: A laser beam is
incident from below through a transparent medium, called medium 1, and
hits the plane, horizontal surface towards the transparent medium 2 at
an angle of incidence, $\theta_1$, which is greater than the critical
angle, $\theta_{crit}=\arcsin(n_2/n_1)$, for total reflection. This
establishes an evanescent field above the plane surface, which, in
turn, is regarded as the incident field falling upon a compact sphere
of radius $a$ and complex refractive index $\bar{n}_3$. The sphere is
centered at the origin with the $x$ axis pointing vertically
upwards. If $\bar{\epsilon}_3$ is the complex permittivity of the
(non-magnetic) sphere, we will require it to be related to the
effective conductivity, $\sigma$, through the equation
\begin{equation}
\bar{\epsilon}_3=\epsilon_3+i \sigma /\omega, \label{2}
\end{equation}
hence $\bar{n}_3=\sqrt{\bar{\epsilon}_3/\epsilon_0}$.

Here, a remark is called for as regards our formal use of the
dispersion relation.  For simplicity, taking the medium to have only
one resonance frequency, we can write the complex permittivity as
\begin{equation}
\bar{\epsilon}(\omega)=\epsilon_3 +i\frac{\epsilon_0
\omega_p^2}{\omega(\gamma-i\omega)};\label{epsilon}
\end{equation}
cf., for instance, Eq.\ (7.56) in Jackson \cite{jackson99}. Here the
first term $\epsilon_3$ describes the contribution from the dipoles
and can be taken to be constant, whereas the second term describes the
dispersive properties of the free electrons.  $\gamma$ is the damping
constant. Comparison between Eqs.\ (\ref{2}) and (\ref{epsilon}) gives
for the frequency dependent term $\sigma(\omega)$
\begin{equation}
\sigma(\omega)=\frac{\epsilon_0\omega_p^2}{\gamma-i\omega}.
\end{equation}
Typically $\gamma \sim 10^{12}\, {\rm s}^{-1}$, so that we can
identify the frequency dependent quantity $\sigma=\sigma(\omega)$ with
a constant, equal to the static conductivity $\sigma_{\rm stat}$, only
for frequencies $\nu=\omega/2\pi$ less than about 10 GHz. Our
frequencies are obviously higher: a vacuum wavelength of
$\lambda_0=1.06\, \mu$m corresponds to $\omega_0=1.78\times
10^{15}\,\rm{s}^{-1}$. Thus our $\sigma(\omega)$ is an {\it effective}
conductivity, defined in accordance with the low-frequency relation
(2), but no longer identifiable with the static conductivity
$\sigma_{\rm stat}$. The actual value of $\sigma(\omega)$ has to be
evaluated in each case.

Take water as an example, for which $\bar{n}=1.33+5.0\times
10^{-6}\,i$ for the actual wavelength \cite{barton88}. Writing in
general the complex refractive index as $\bar{n}=n'+in''$ we get, when
$\sigma/\epsilon_0\omega \ll 1$, $\sigma/\epsilon_0\omega =2n'n''$,
which in our case amounts to only $1.33\times 10^{-5}$, corresponding
to
\begin{equation}
\sigma=0.21\, {\rm S/m}.
\end{equation}
This can be compared with the static conductivity for pure water,
which is much smaller, $\sigma_{\rm stat}=2\times 10^{-4}$ S/m
\cite{stratton41}. It is notable that the value of $\sigma$ is real.

Increasing the incident wavelength somewhat, the absorptive effect can
be enhanced. Thus, if we consider glass, and choose
$\lambda_0=8.0\,\mu$m, we have $\bar{n}=0.411+0.323\, i$
\cite{philipp85}.  With $\omega=2.36\times 10^{14}\, {\rm s}^{-1}$ in
this case we get for the effective conductivity quite an appreciable
value:
\begin{equation}
\sigma=550 \, {\rm S/m}.
\end{equation}
(Actually the parameter $\sigma/\epsilon_0\omega$ is as high as
0.265 in this case, and can hardly be assumed to be a "small"
parameter any longer if $\lambda_0$ is increased further.)

We assume now that the incoming beam in medium 1 is a plane wave, such
that its electric and magnetic field amplitudes are related through
$H_0=\sqrt{\epsilon_0/\mu_0}\, E_0$. We introduce the non-dimensional
wave number, $\alpha$, in medium 2, and the relative radial coordinate
$\tilde{r}$:
\begin{equation}
\alpha=k_2a=n_2\omega a/c,~~~~\tilde{r}=r/a,\label{3}
\end{equation}
and make use of the standard spherical coordinates
$(r,\theta,\varphi)$ centered at the origin of the sphere. Expansions
for the incident electric field ${\bf E}^{(i)} = (E_r^{(i)},
E_\theta^{(i)}, E_\varphi^{(i)})$, the scattered field ${\bf E}^{(s)}
= (E_r^{(s)}, E_\theta^{(s)}, E_\varphi^{(s)})$, and the internal
field in the sphere, ${\bf E}^{(w)} = (E_r^{(w)}, E_\theta^{(w)},
E_\varphi^{(w)})$, are for convenience collected in Appendix A
together with the analogous expansions of the magnetic field. In these
expressions, we have imposed the electromagnetic boundary conditions
at the spherical surface, $r=a$.  We will need use of the
Riccati-Bessel functions $\psi_l$ and $\xi_l^{(1)}$:
\begin{equation}
\psi_l(x)=xj_l(x),~~~~\xi_l^{(1)}(x)=xh_l^{(1)}(x).\label{4}
\end{equation}
Here, $j_l$ and $h_l$ are the spherical Bessel and Hankel functions,
satisfying the Wronskian $W\{\psi_l, \xi_l^{(1)}\}=i$. There are three
groups of expansion coefficients: $\{A_{lm}, B_{lm}\}$ for the
incident field, $\{a_{lm},b_{lm}\}$ for the scattered field, and
$\{c_{lm},d_{lm}\}$ for the internal field. As for the first-mentioned
coefficients, they are determined by inverting the formulas (\ref{a1})
and (\ref{a4}) for $E_r^{(i)}$ and $H_r^{(i)}$:
\begin{eqnarray}
A_{lm}\!\! &=&\!\! \frac{(b/a)^2}{E_0 l(l+1)\psi_l(k_2b)}\!\int_\Omega\!\!\! E_r^{(i)}(b,
	\theta,\varphi) Y_{lm}^*(\theta,\varphi)d\Omega, \label{5} \\
B_{lm}\!\! &=&\!\! \frac{(b/a)^2}{H_0 l(l+1)\psi_l(k_2b)}\!\int_\Omega\!\!\! H_r^{(i)}
	(b,\theta,\varphi) Y_{lm}^*(\theta, \varphi)d \Omega,~~~~ \label{6}
\end{eqnarray}
where $d\Omega=\sin \theta \,d\theta \,d\varphi$. Here, the angular
integrations are taken over a sphere of arbitrary radius $b$. This
freedom of choice for $b$ gives us a calculational advantage. The
results for $A_{lm}$ and $B_{lm}$ are only related to the incident
field, and they have to be independent of $b$. Calculational
insensitivity with respect to various input values for $b$ thus serves
as a check of the evaluations of $A_{lm}$ and $B_{lm}$. In the
following, we will suppress the writing of the time factors $\exp(-i
\omega t)$.

Our expansion procedure follows the conventions of Barton {\it et \
al.} \cite{barton88,barton89}. This procedure was also followed in
Ref.\ \cite{almaas95}. In the case where $b$ is chosen equal to $a$,
Eqs.\ (\ref{5}) and (\ref{6}) agree with Eqs.\ (A.23) and (A.24) of
Ref.\ \cite{barton89}.

In order to calculate $A_{lm}$ and $B_{lm}$, it is necessary to know
the incident field components $E_r^{(i)} \equiv E_r^{(2)}$ and
$H_r^{(i)} \equiv H_r^{(2)}$ in region 2. We introduce the parameters
\begin{equation}
\!\beta=\frac{n_1\omega}{c}\sqrt{\sin^2 \theta_1-n_{21}^2},
   \gamma=\frac{n_1\omega}{c}\sin \theta_1,
   n_{21}=\frac{n_2}{n_1},\label{7}
\end{equation}
and take the following expressions for the amplitude ratios,
$T_\parallel = E_\parallel^{(2)} / E_\parallel^{(1)}$, and $T_\perp =
E_\perp^{(2)} / E_\perp^{(1)}$ at the surface of the substrate into
account:
\begin{eqnarray}
T_\parallel &=& \frac{2n_{21}\cos \theta_1}{n_{21}^2\cos
   \theta_1+i\sqrt{\sin^2\theta_1-n_{21}^2}},\label{8}
\end{eqnarray}
\pagebreak
\begin{eqnarray}
T_\perp &=& \frac{2 \cos \theta_1}{\cos \theta_1 + i
   \sqrt{\sin^2\theta_1-n_{21}^2}}. \label{9}
\end{eqnarray}
Here, $E_\parallel$ refers to the field component in the plane of
incidence ($p$ polarization), and $E_\perp$ to the field orthogonal to
it ($s$ polarization). For the radial field components, we find
\begin{widetext}
\begin{eqnarray}
\!E_r^{(i)} &=&\!\! \Bigg\{ \frac{1}{n_{21}}T_ \parallel E_\parallel^{(1)}
       [\sin \theta_1\sin \theta \cos \varphi
       -i\sqrt{\sin^2\theta_1-n_{21}^2}\cos \theta]
       + T_\perp E_\perp^{(1)} \sin \theta \sin \varphi \Bigg\} \times
       \exp[-\beta (x+h)+i\gamma z],\label{10} \\
\!H_r^{(i)} &=&\!\! \Bigg\{T_\perp H_\parallel^{(1)}[-\sin \theta_1 \sin
       \theta \cos \varphi+i\sqrt{\sin^2 \theta_1-n_{21}^2} 
       \cos \theta]
       +n_{21}T_\parallel H_\perp^{(1)}\sin\theta\sin\varphi \Bigg\}
       \times \exp[-\beta (x+h)+i\gamma z],\label{11}
\end{eqnarray}
\end{widetext}
where
\begin{equation}
H_\parallel^{(2)}/H_\parallel^{(1)} = n_{21}T_\perp,~~~~H_\perp^{(2)} /
   H_\perp^{(1)} = n_{21}T_\parallel, \label{12}
\end{equation}
at the surface of the substrate. (Readers interested in background
material for our developments can {\it e.g.}~ consult the articles
\cite{almaas95,barton88,barton89,farsund96}, as well as the standard
texts \cite{stratton41,kerker69,born91}.) Upon insertion into Eqs.\
(\ref{5}) and (\ref{6}), we find that the coefficients $A_{lm}$ and
$B_{lm}$ for the case of $s$ ($\perp$) and $p$ ($\parallel$)
polarization are related through
\begin{eqnarray}
A_{lm}^\perp &=& \frac{T_\perp}{n_2T_\parallel}B_{lm}^\parallel, \label{13}\\
B_{lm}^\perp &=& -\frac{n_2T_\perp}{T_\parallel}A_{lm}^\parallel,\label{14}
\end{eqnarray}
hence, it is sufficient to only consider one polarization direction in
the calculations of the expansion coefficients, and in the following,
we will choose the case of $p$ polarization. By inserting the
expressions for the incident field into Eqs.\ (\ref{5}) and (\ref{6}),
we obtain the following expressions for the expansion coefficients:
\begin{eqnarray}
A_{lm}^\parallel &=& \frac{\alpha_1(l,m)}{n_{21}} T_\parallel e^{-\beta
	  h}\bigg[\sin\theta_1 Q_1(l,m) \nonumber \\
	&& - i\sqrt{\sin^2\theta_1-n_{21}^2}\,Q_2(l,m) \bigg],\label{15} \\
B_{lm}^\parallel &=& n_2\alpha_1(l,m)T_\parallel\,e^{-\beta h}\,Q_3(l,m),
  \label{16}
\end{eqnarray}
where we have defined
\begin{eqnarray}
\alpha_1(l,m) &=& \left[ \frac{2l+1}{4\pi} \frac{(l-m)!}{(l+m)!}
   \right]^{1/2} \frac{(b/a)^2}{l(l+1)\psi_l(k_2b)}, \label{17} \\
Q_1(l,m) &=& 2\pi (-1)^{m-1}\int_0^{\pi /2}\sin^2\theta 
     \left\{ \begin{array}{c} \cos \\
                              i\sin
             \end{array}
     \right\} (\gamma b\cos\theta) \nonumber \\ && 
\!\!\!\!\times P_l^m(\cos\theta) \left[ I_{|m-1|}(u) + I_{|m+1|}(u)\right]d\theta, \label{18} 
\end{eqnarray}
\begin{eqnarray}
Q_2(l,m) &=& 4\pi (-1)^m \int_0^{\pi /2} \sin \theta \cos \theta 
	\left\{ \begin{array}{c} i\sin \\
                                  \cos 
                \end{array}
        \right\} (\gamma b\cos\theta)\, \nonumber \\
	&&  \times P_l^m(\cos \theta)  I_{|m|}(u)d\theta,\label{19} \\
Q_3(l,m) &=& 4\pi i(-1)^m \frac{m}{\beta b} \int_0^{\pi /2} \sin\theta 
         \left\{ \begin{array}{c}\cos \\
                                 i\sin
                 \end{array}
         \right\} (\gamma b\cos \theta) \nonumber \\ 
	&& \times P_l^m (\cos \theta) I_{|m|}(u) d \theta,\label{20}
\end{eqnarray}
where $u=\beta b\sin\theta$, 
$l+m$ is $ \left\{ \begin{array}{c}\mathrm{even} \\
                         \mathrm{odd}
                         \end{array}
                 \right\}$, and 
$I_m(z)=i^{-m}J_m(iz)$ is a modified Bessel function. (In Ref.\
\cite{almaas95}, Eqs.\ (46) and (49), were missing factors of 2.)
Again, the calculated values of $A_{lm}^\parallel$ and
$B_{lm}^\parallel$ have to be numerically independent of the value
chosen for the radius $b$.

Once $A_{lm}$ and $B_{lm}$ are known, the other sets of coefficients,
$\{a_{lm},b_{lm}\}$ and $\{c_{lm}, d_{lm}\}$, can readily be
calculated using Eqs.\ (\ref{a19})--(\ref{a22}). In appendix B, we
give a table of $A_{lm}$ and $B_{lm}$ up to $l_{\mathrm{max}} = 7$ for
$p$ polarization.

\section{The surface force}

Let us, as indicated above, write the total radiation force ${\bf F}$
on the sphere as a sum of two contributions:
\begin{equation}
{\bf F} = {\bf F}^{\mathrm{surf}} + {\bf F}^{\mathrm{abs}}, \label{21}
\end{equation}
where ${\bf F}^{\mathrm{surf}}$ is the result of the force density
${\bf f}$ acting in the surface layer, {\it i.e.} Eq.\ (\ref{1}), and
${\bf F}^{\mathrm{abs}}$ is caused by the absorption in the sphere. By
integrating across the surface layer, the components
$F_i^{\mathrm{surf}}$ of ${\bf F}^{\mathrm{surf}}$ can be related to
the surface integral of Maxwell's stress tensor $S_{ik}$ on the
outside of the sphere:
\begin{equation}
F_i^{\mathrm{surf}} = -\int S_{ik} n_k dS,\label{22}
\end{equation}
${\bf n}$ being the outward normal (cf. for instance, Refs.\
\cite{brevik79,barton89}). [This expression assumes that the effect
from absorption is negligible, in other words that the size of the
particle is much smaller than the skin depth $\delta=(2/\mu_0 \omega
\sigma)^{1/2}$.] If $a\ll \lambda_0$, and the susceptibility of the
particle is small, the formula for the total force ${\bf F}$ goes over
to the conventional expression for dilute particles:
\begin{equation}
{\bf F}=6\pi
a^3\frac{\epsilon''}{(\epsilon+2\epsilon_0)^2}\epsilon_0{\bf
k}|E^{(i)}|^2;
\end{equation}
cf., for instance, Refs.~\cite{chaumet00,ashkin86,landau84}.

It should be stressed that we are calculating the force on the sphere
in two steps: First, the surface integral (26) is evaluated assuming
the refractive index to be real. That means, the coefficients $A_{lm}$
and $B_{lm}$ below are calculated on the basis of a real
$n_3$. Thereafter, the absorptive part ${\bf F}^{\mathrm{abs}}$is
calculated separately, via the force density ${\bf J\times B}$.

Using the general expressions for the incident and the scattered
field, Eqs.\ (\ref{a1}) -- (\ref{a12}), we can re-write this surface
integral in terms of the complex expansion coefficients $A_{lm}$,
$B_{lm}$, $a_{lm}$ and $b_{lm}$:
\begin{widetext}
\begin{eqnarray}
\frac{F_x^{\mathrm{surf}} + i F_y^{\mathrm{surf}}}{\epsilon_0 E_0^2 a^2} &=&
    \frac{i\alpha^2}{4}\sum_{l=1}^\infty \sum_{m=-l}^l \Bigg\{
    \left[\frac{(l+m+2)(l+m+1)}{(2l+1)(2l+3)}\right]^{1/2}l(l+2)
    \bigg( 2n_2^2 a_{lm} a_{l+1,m+1}^* + n_2^2 a_{lm} A_{l+1,m+1}^* \nonumber\\
    &+& n_2^2 A_{lm} a_{l+1,m+1}^* 
    + 2 b_{lm} b_{l+1,m+1}^* + b_{lm} B_{l+1,m+1}^*
    + B_{lm}b_{l+1,m+1}^* \bigg)
    + \left[ \frac{(l-m+1)(l-m+2)}{(2l+1) (2l + 3)} \right]^{1/2} \nonumber\\
    &\times& l(l+2)\bigg(2n_2^2 a_{l+1,m-1} a_{lm}^* +n_2^2 a_{l+1,m-1}A_{lm}^*
    + n_2^2 A_{l+1,m-1} a_{lm}^* + 2 b_{l+1,m-1} b_{lm}^* +b_{l+1,m-1}B_{lm}^*
    \nonumber \\ &+& B_{l+1,m-1}b_{lm}^* \bigg)
    -\bigg[(l+m+1)(l-m)\bigg]^{1/2} n_2 \bigg( -2 a_{lm} b_{l,m+1}^* + 2 b_{lm}
    a_{l,m+1}^* -a_{lm} B_{l,m+1}^* \nonumber \\ 
    &+& b_{lm} A_{l,m+1}^* + B_{lm} a_{l,m+1}^* 
     -  A_{lm} b_{l,m+1}^* \bigg) \Bigg\}, \label{23} 
\end{eqnarray}
\begin{eqnarray}
\frac{F_z^{\mathrm{surf}}}{\epsilon_0 E_0^2a^2} &=& -\frac{\alpha^2}{2} 
    \sum_{l=1}^\infty \sum_{m=-l}^l \left[ \frac{(l-m+1)(l+m+1)}{(2l+1)
    (2l+3)}\right]^{1/2} l(l+2) 
    \Im \Big[ 2 n_2^2 a_{l+1,m} a_{lm}^* + n_2^2 a_{l+1,m} A_{lm}^* 
    +n_2^2A_{l+1,m}a_{lm}^* \nonumber \\ 
    &+& 2 b_{l+1,m} b_{lm}^* + b_{l+1,m} B_{lm}^* + B_{l+1,m} b_{lm}^* 
    + n_2 m \big( 2 a_{lm} b_{lm}^* + a_{lm} B_{lm}^* +A_{lm} b_{lm}^* 
    \big) \Big]. \label{24}
\end{eqnarray}
\end{widetext}
We shall not dwell much on these expressions, since our main interest,
as mentioned above, is the vertical force $F_x^{\mathrm{abs}}$
associated with the absorption. Consider, however, the horizontal
surface force component $F_z^{\mathrm{surf}}$. With an incident beam
power (in vacuum) of $P = 150$ mW, distributed over a circular
cross-sectional area of diameter 10 $\mu$m, we find the magnitude of
the Poynting vector to be $(c/2) \epsilon_0 E_0^2 = 19.0
~MW/m^2$. Taking the radius of the sphere $a = 1~\mu$m, we find the
denominator in Eq.\ (\ref{24}) to be $\epsilon_0 E_0^2 a^2 = 0.13$
pN. If the sphere is made of glass ($n_3 = 1.50$) and the surrounding
medium 2 is water ($n_2 = 1.33$), we then have $\alpha = 2\pi
a/\lambda_2 = 7.9$, since the wavelength in water for Nd:YAG laser
light is $\lambda_2 = 1.06/ 1.33 = 0.80~\mu$m. Using Fig. 5 in
Ref.~\cite{almaas95}, we find that $F_z^{\mathrm{surf}}$ is positive,
as it should be, and is approximately equal to $0.010$ pN in the case
of $p$ polarization. Since the Reynolds number is very low, the Stokes
drag formula $D = 6\pi \mu av$ is applicable, using $\mu = 1.0 \times
10^{-3}$ Pa s as the dynamic viscosity of water. Putting
$F_z^{\mathrm{surf}} = D$, we obtain the drift velocity of the sphere
to be $v = 0.53~ \mu$ m/s. This is an order of magnitude agreement
with the observations of Kawata and Sugiura \cite{kawata92}; from
their Fig.\ $4$, one infers that $v \sim 1-2~ \mu$m/s. This 
agreement is satisfactory;  there is no need to
carry out a complicated calculation to find the absorptive correction
to the horizontal radiation force. In the following, we focus
 attention on the vertical absorptive part of
the force.

\section{The vertical absorptive force}

To calculate the absorptive force, we first replace Eq.\ (\ref{1}) by
the general expression
\begin{equation}
{\bf f} = \rho {\bf E} + {\bf J \times B} - \frac{1}{2} E^2 {\bf \nabla }
   \epsilon, \label{25}
\end{equation}
where $\rho$ and ${\bf J}$ are the external charge and current
densities in the medium. A general derivation of this force expression
can be found, for instance, in Stratton's book
\cite{stratton41}. Also, in the review article of one of the present
authors \cite{brevik79}, it was found that the force expression is
able to describe the outcome of known experiments.  We assume no
external charges to be present, so that $\rho =0$. Writing ${\bf J} =
\sigma {\bf E}$ where $\sigma$ is the frequency-dependent
conductivity, assumed to be real (cf. Eq.\ (5)), we get in
complex representation the following expression for the absorptive
force:
\begin{equation}
{\bf F}^{\mathrm{abs}} = \frac{1}{2} \Re \int {\bf J \times B^* } dV = 
   \frac{\sigma \mu_0}{2} \Re \int {\bf E \times H^*}\ dV, \label{26}
\end{equation}
where the integral is evaluated over the volume of the sphere.

In cases of current interest, we can simplify our calculation: In view
of the assumed smallness of $\sigma/\epsilon_0\omega$, we replace the
fields inside the sphere with those calculated when assuming the
material to be a perfect dielectric. That is, we can replace the
complex refractive index $\bar{n}_3$ by its real part $n_3$. This
facilitates the calculation of the volume integrals. The expressions
for the internal fields are given by Eqs.\ (\ref{a13})--(\ref{a18}),
in spherical coordinates. For notational convenience, we will omit the
superscripts $(w)$.

With $\tilde{r}=r/a$ we can write the vertical absorptive force as
\begin{eqnarray}
F_x^{\mathrm{abs}}&=&\frac{\sigma \mu_0}{2}a^3\,\Re \int_0^{2\pi} \!\!\!\!
    d\varphi \int_0^\pi\!\!\! \sin\theta d\theta 
    \int_0^1\!\!\!\tilde{r}^2 d\tilde{r}
    \Bigg[ ({\bf E\times H^*)}_r\nonumber\\
    \times\sin\theta\!\!\!\!\!\!\!&&\!\!\!\!\cos\varphi +\!
    ({\bf E \times H^*)}_\theta \cos\theta \cos \varphi -\!  {\bf (E\times
    H^*)}_\varphi \sin\varphi \Bigg], \nonumber \\
    &=&\frac{\sigma \mu_0}{2}a^3\,\Re \sum_{i=1}^6I_i, \label{27}
\end{eqnarray}
where we have defined
\begin{eqnarray}
I_1 &=& \int_0^{2\pi}\!\! \cos \varphi \, d\varphi \int_0^\pi\!\! \sin^2 \theta
        d\theta \int_0^1\!\! \tilde{r}^2 d\tilde{r} E_\theta H_\varphi^*, 
	\label{28}
\end{eqnarray}
\begin{eqnarray}
I_2 &=& -\int_0^{2\pi}\!\! \cos \varphi\, d\varphi\int_0^\pi\!\! \sin^2 \theta 
	d\theta \int_0^1\!\!\tilde{r}^2 d\tilde{r} E_\varphi H_\theta^*, 
	\label{29}\\
I_3 &=& \int_0^{2\pi}\!\!\!\!\cos\varphi \, d\varphi\int_0^\pi\!\!\!\cos\theta 
	\sin \theta d\theta \int_0^1\!\! \tilde{r}^2 d\tilde{r}E_\varphi H_r^*,
	\label{30}\\
I_4 &=&\!\! -\int_0^{2\pi}\!\!\!\!\!\!\cos\varphi \, d\varphi\int_0^\pi\!\!\!\!
        \cos\theta 
	\sin \theta d\theta \int_0^1\!\! \tilde{r}^2 d\tilde{r}E_r H_\varphi^*,
	\label{31}\\
I_5 &=& -\int_0^{2\pi}\!\! \sin \varphi \, d\varphi \int_0^\pi\!\! \sin \theta 
	d\theta \int_0^1\!\! \tilde{r}^2d\tilde{r} E_r H_\theta^* ,\label{32}\\
I_6 &=& \int_0^{2\pi}\!\! \sin \varphi\, d\varphi \int_0^\pi\!\! \sin \theta 
	d\theta \int_0^1\!\! \tilde{r}^2 d\tilde{r} E_\theta H_r^*.\label{33}
\end{eqnarray}
It is now convenient to introduce the operators
\begin{eqnarray}
\mathcal{L} &=& \sum_{l=1}^\infty \sum_{j=1}^\infty \int_0^{n_{32}\alpha} 
	du,\label{34} \\
\mathcal{M}_{lj}^{( \pm )} &=& \sum_{m=-l}^l \sum_{k=-j}^j \left( 
	\delta_{m,k-1} \pm \delta_{m,k+1} \right). \label{35}
\end{eqnarray}
We write the two first of the $I_i$'s as a sum of four terms:
\begin{equation}
I_i = i \pi \alpha n_1 c \epsilon_0 E_0^2 \sum_{k=1}^4 I_{ik},
       ~~~~i=1,2. \label{36}
\end{equation}
and remaining $I_i$'s, which are simpler since they are each composed of
only two terms:
\begin{equation}
I_i = i \pi \alpha n_1 c \epsilon_0 E_0^2 \sum_{k=1}^2 I_{ik},
	~~~~i=3,4,5,6. \label{46}
\end{equation}
where (in the following, $l$, $m$, $j$ and $k$ denote summation indices as
defined in Eqs. (\ref{34}) and (\ref{35}))
\begin{widetext}
\begin{eqnarray}
I_{11} &=& -n_{32} \mathcal{L} \Bigg[ \psi_l'(u)
       \psi_j'(u) \mathcal{M}_{lj}^{(+)} \Bigg\{ kc_{lm} d_{jk}^* 
       C_{lm} C_{jk} {\mathcal R}_1(l,m,j,k) \Bigg\} \Bigg], \label{37}\\
I_{12} &=& -n_{32}^2 n_2 \mathcal{L} \Bigg[ \psi_l'(u)
       \psi_j(u) \mathcal{M}_{lj}^{(+)} \Bigg\{c_{lm} c_{jk}^*
       C_{lm} C_{jk} {\mathcal R}_2 (l,m,j,k) \Bigg\} \Bigg], \label{38} \\
I_{13} &=& \frac{1}{n_2} \mathcal{L} \Bigg[ \psi_l(u)
       \psi_j'(u) \mathcal{M}_{lj}^{(+)} \Bigg\{ m k d_{lm}
       d_{jk}^* C_{lm} C_{jk} {\mathcal R}_3(l,m,j,k)
	\Bigg\} \Bigg], \label{39} \\
I_{14} &=&n_{32}  \mathcal{L} \Bigg[ \psi_l(u) \psi_j(u)
       \mathcal{M}_{lj}^{(+)} \Bigg\{ md_{lm} c_{jk}^* C_{lm}
       C_{jk} {\mathcal R}_1(j,k,l,m) \Bigg\} \Bigg], \label{40}\\
I_{21} &=& -n_{32} \mathcal{L} \Bigg[ \psi_l'(u)
       \psi_j'(u) \mathcal{M}_{lj}^{(+)} \Bigg\{ m c_{lm} d_{jk}^*
       C_{lm} C_{jk} {\mathcal R}_1 (j,k,l,m) \Bigg\} \Bigg], \label{41}\\
I_{22} &=& -n_{32}^2 \mathcal{L}\Bigg[ \psi_l'(u)
       \psi_j(u) \mathcal{M}_{lj}^{(+)} \Bigg\{ m k c_{lm} c_{jk}^*
       C_{lm} C_{jk} {\mathcal R}_3 (l,m,j,k)\Bigg\}\Bigg], \label{42}
\end{eqnarray}
\begin{eqnarray}
I_{23} &=& \frac{1}{n_2} \mathcal{L} \Bigg[ \psi_l(u)
       \psi_j'(u) \mathcal{M}_{lj}^{(+)} \Bigg\{ d_{lm} d_{jk}^*
       C_{lm} C_{jk} {\mathcal R}_2(l,m,j,k) \Bigg\} \Bigg], \label{43}\\
I_{24} &=& n_{32} \mathcal{L}\Bigg[ \psi_l(u) \psi_j(u)
       \mathcal{M}_{lj}^{(+)} \Bigg\{ kd_{lm} c_{jk}^* C_{lm}
       C_{jk} {\mathcal R}_1(l,m,j,k) \Bigg\} \Bigg], \label{44}\\
I_{31} &=& n_{32} \mathcal{L}\Bigg[ \frac{\psi_l'(u)}{u}
       \psi_j(u)j(j+1) \mathcal{M}_{lj}^{(+)} \Bigg\{ mc_{lm}
       d_{jk}^* C_{lm} C_{jk}  {\mathcal R}_4(l,m,j,k) \Bigg\} \Bigg], 
       \label{47} \\
I_{32} &=& -\frac{1}{n_2} \mathcal{L} \Bigg[ \frac{
       \psi_l(u)}{u} \psi_j(u) j (j + 1) \mathcal{M}_{lj}^{(+)} \Bigg\{
       d_{lm} d_{jk}^* C_{lm} C_{jk}  {\mathcal R}_5 (l,m,j,k) \Bigg\} 
       \Bigg], \label{48} \\
I_{41} &=& n_{32} \mathcal{L} \Bigg[ \frac{\psi_l(u)}{u}
       \psi_j'(u) l(l + 1) \mathcal{M}_{lj}^{(+)} \Bigg\{ kc_{lm}
       d_{jk}^* C_{lm} C_{jk} {\mathcal R}_4 (l,m,j,k) \Bigg\} 
       \Bigg], \label{49}\\
I_{42} &=& n_{32}^2 n_2 \mathcal{L} \Bigg[ \frac{
       \psi_l(u)}{u} \psi_j(u) l (l + 1) \mathcal{M}_{lj}^{(+)} \Bigg\{
       c_{lm} c_{jk}^* C_{lm} C_{jk} {\mathcal R}_5 (j,k,l,m) \Bigg\}
       \Bigg], \label{50}\\
I_{51} &=& -n_{32} \mathcal{L} \Bigg[ \frac{
       \psi_l(u)}{u} \psi_j'(u) l(l + 1) \mathcal{M}_{lj}^{(-)} \Bigg\{
       c_{lm} d_{jk}^* C_{lm} C_{jk} {\mathcal R}_1(j,k,l,m) \Bigg\} \Bigg], 
       \label{51} \\
I_{52} &=& - n_{32}^2 n_2  \mathcal{L} \Bigg[ \frac{
       \psi_l(u)}{u} \psi_j(u) l(l + 1) \mathcal{M}_{lj}^{(-)}
       \Bigg\{k c_{lm} c_{jk}^* C_{lm} C_{jk} {\mathcal R}_3 (l,m,j,k)
       \Bigg\}\Bigg], \label{52}\\
I_{61} &=& n_{32} \mathcal{L} \Bigg[ \frac{
       \psi_l'(u)}{u} \psi_j(u) j(j + 1) \mathcal{M}_{lj}^{(-)} \Bigg\{
       c_{lm} d_{jk}^*  C_{lm} C_{jk} {\mathcal R}_1 (l,m,j,k) \Bigg\} \Bigg],
       \label{53} \\
I_{62} &=& - \frac{1}{n_2} \mathcal{L} \Bigg[ \frac{
        \psi_l(u)}{u} \psi_j(u) j(j + 1) \mathcal{M}_{lj}^{(-)} \Bigg\{ m
        d_{lm} d_{jk}^* C_{lm} C_{jk} {\mathcal R}_3(l,m,j,k)
	\Bigg\} \Bigg], \label{54}
\end{eqnarray}
\end{widetext}
with
\begin{eqnarray}
{\mathcal R}_1(l,m,j,k)\!\! &=&\!\! \int_0^\pi\!\!\! \sin\theta~ \frac{d P_l^m 
	\left(\cos\theta\right)}{d \theta}\,  P_j^k \left(\cos\theta\right) 
        d \theta, \label{r1} \\
{\mathcal R}_2(l,m,j,k)\!\! &=&\!\!\int_0^\pi\!\!\!\sin^2\theta\,\frac{d P_l^m 
	\left(\cos\theta\right)}{d \theta}\, \frac{d P_j^k 
	\left(\cos\theta\right)}{d \theta}\, d \theta,~~~~~~\label{r2}\\
{\mathcal R}_3(l,m,j,k)\!\! &=&\!\!\! \int_0^\pi\!\!\! P_l^m 
	\left(\cos\theta\right)
	\, P_j^k \left(\cos\theta\right)\, d \theta, \label{r3}\\
{\mathcal R}_4(l,m,j,k)\!\! &=&\!\!\! \int_0^\pi\!\!\! \cos\theta\,  P_l^m 
	\left(\cos\theta\right)\, P_j^k \left(\cos\theta\right)
	\, d\theta,\label{r4} \\
{\mathcal R}_5(l,m,j,k)\!\!\! &=&\!\!\!\int_0^\pi\!\!\!\!\cos\theta \sin\theta 
        \frac{d P_l^m\! \left(\cos\theta\right)}{d \theta}\,
	 P_j^k\! \left(\cos\theta\right) d\theta,~~~~\label{r5} 
\end{eqnarray}
and the definition
\begin{equation}
C_{lm} = \sqrt{\frac{(2l+1)(l-m)!}{4\pi (l+m)!}}. \label{45}
\end{equation}
Altogether, writing the non-dimensional vertical absorptive force as
$Q_x^{\mathrm{abs}}$, where
\begin{equation}
Q_x^{\mathrm{abs}}=\frac{F_x^{\mathrm{abs}}}{\epsilon_0 E_0^2 a^2}, \label{55}
\end{equation}
we get
\begin{equation}
\!\!Q_x^{\mathrm{abs}}\!\!= -\! \sigma \mu_0 c \lambda  \frac{n_{12}\alpha^2}{4} 
     ~\Im \left[\sum_{i=1}^2\sum_{k=1}^4 I_{ik}+\!\sum_{i=3}^6
     \sum_{k=1}^2I_{ik} \right]. \label{56}
\end{equation}
Note that the prefactor in the above equations, as well as the terms
$I_{ik}$, are non-dimensional. The expression (\ref{56}) is our main
result.

\section{Results}

\subsection{Numerics}

\begin{figure}[!t]
\includegraphics[width=8.5cm]{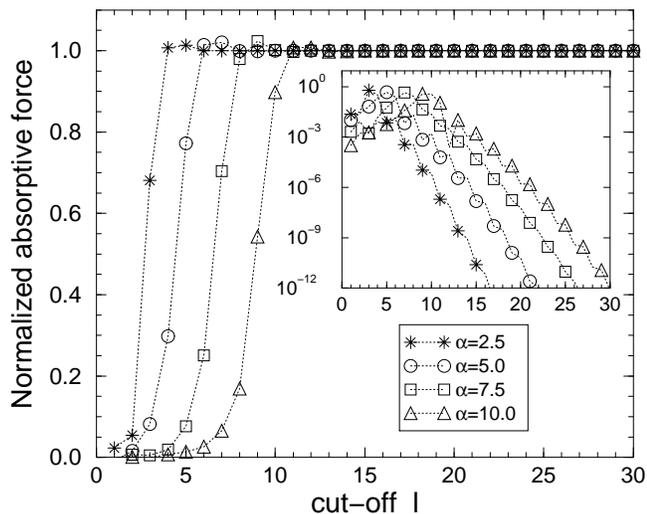}
\caption{Plot of the total absorptive force versus increasing cut-off
         in Eq.\ (\ref{34}) for 4 values of $\alpha$: $2.5$, $5.0$, $7.5$,
         and $10.0$, with $n_1 = 1.75$, $n_2 = 1.33$, and $n_3 =
         1.50$. The inset shows the absolute value of the difference
         in the total force using $l$ and $(l-1)$ respectively, as
         cut-offs.}
\label{fig:2}
\end{figure}

We implemented an adaptive Gauss-Kronrod rule to calculate the
integrals in Eqs.\ (\ref{18}) -- (\ref{20}) and (\ref{37}) --
(\ref{54}). By employing the symmetry properties of the integrals over
the Legendre polynomials (Eqs.\ (\ref{r1}) -- (\ref{r5})), we were
able to greatly reduce the number of integrals to be computed.  To
make sure that the sum over $l$ and $j$ in Eq.\ (\ref{34}) converged
properly for our chosen cut-off value, we plotted the total absorptive
force on the sphere versus increasing cut-off, $l_{\mathrm{max}}$, in
$l$ (we used the same cut-off for $l$ and $j$). In Fig.\ \ref{fig:2},
we give an example for $p$ polarization using several values of
$\alpha$. The total force is normalized to the $l_{\mathrm{max}}=30$
result; all other parameters are given in the figure caption.  This
figure clearly suggests that, for a given value of $\alpha$, there is
a narrow range in $l$ which accounts for most of the absorptive force.
Moreover, it is seen that this range of important $l$-values moves to
higher $l$ and broadens with increasing $\alpha$.  In the inset, we
plot on a linear-log scale the absolute value of the difference in the
total force using $l$ and $(l-1)$ respectively, as cut-offs. The inset
indicates that contributions to the absorptive force from large
$l$-values (large with respect to the $l$-value of the peak) are
exponentially suppressed. Hence, for the range of values for $\alpha$,
$n_1$, $n_2$, and $n_3$ studied in this article, we need only consider
contributions to the force from $l \leq 30$.

\begin{figure}[!t]
\includegraphics[width=8.5cm]{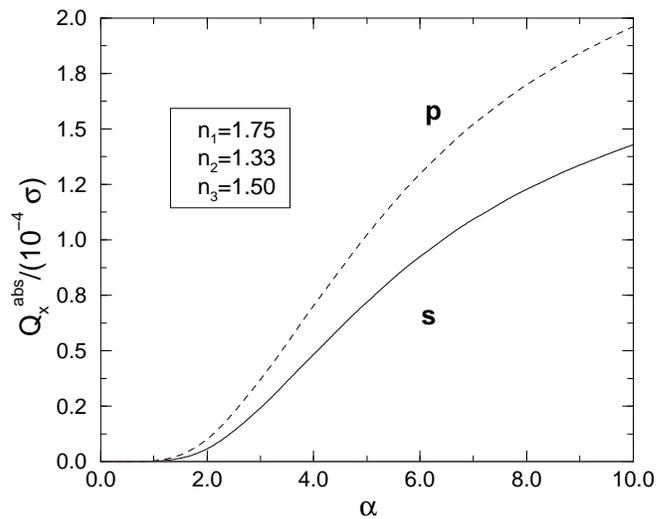}
\caption{Non-dimensional vertical absorptive force $Q_x^{\mathrm{abs}}
         = F_x^{\mathrm{abs}} / ( \epsilon_0 E_0^2 a^2)$ versus
         particle size parameter $\alpha=k_2a$ for the set of
         refractive indices shown. The two states of polarization for
         the incident beam are distinguished.  Here, as well as in the
         subsequent figures, $\theta_1 = 51^\circ$. Also, all figures
         refer to sphere resting on the plate, {\it i.e.} to
         $h=a$. The figure is scaled against the conductivity $\sigma$
         of the sphere.}
\label{fig:3}
\end{figure}

It is of interest to point out that the main features of Fig.\
\ref{fig:2} can be understood from a physical point of view. (This
kind of argument is frequently made use of in connection with
Casimir-related calculations; cf. for instance, Ref.\
\cite{brevik99}). The most significant angular momenta are those that
are of the same order of magnitude as, or are somewhat smaller than,
the angular momentum corresponding to grazing incidence on the
sphere. Consider now a photon with energy $\hbar \omega$ and momentum
$\hbar k_2$ that just touches the surface of the sphere. Its angular
momentum is $\hbar k_2 a$. Setting the angular momentum equal to
$\hbar l$, we thus obtain $l=k_2 a = \alpha$. Thus, the most
significant values of $l$ should be expected to be $l \lesssim
\alpha$. This is seen to be in accordance with the calculated results
in Fig.\ \ref{fig:2}. In view of the crudeness of the argument, the
agreement is actually better than we had reason to expect.

\subsection{Discussion of the figures}

\begin{figure}[!t]
\includegraphics[width=8.5cm]{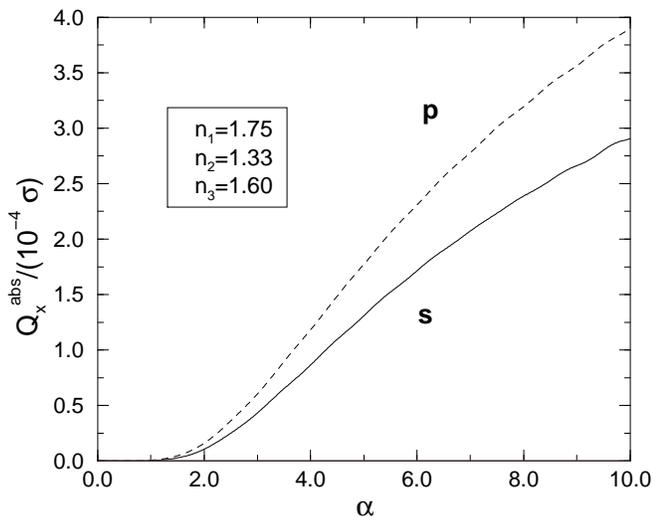}
\caption{Same as for Fig. \ref{fig:3}, but with a higher (dominant
         real part of) refractive index ($n_3=1.60$) in the sphere.}
\label{fig:4}
\end{figure}

\begin{figure}[!t]
\includegraphics[width=8.5cm]{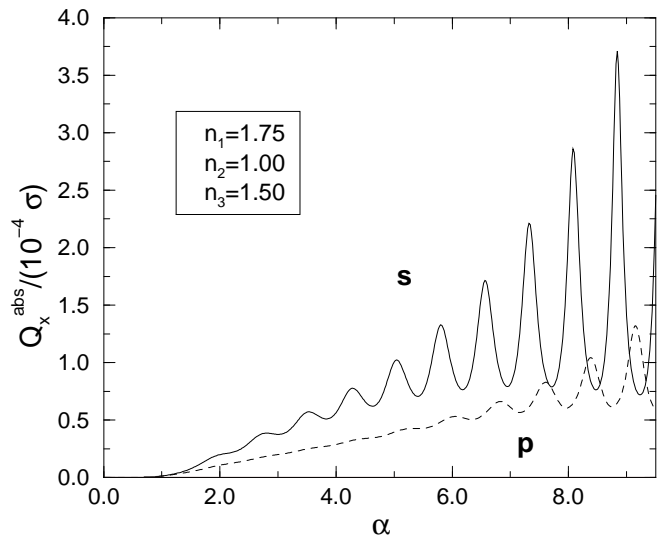}
\caption{Same as for Fig. \ref{fig:3}, but with vacuum ($n_2=1$)
         surrounding the sphere.}
\label{fig:5}
\end{figure}

Illustrative examples of the behavior of the non-dimensional vertical
absorptive force are shown in Figs.\ \ref{fig:3} -- \ref{fig:5}. The
angle of incidence is $\theta_1 = 51^\circ$ everywhere. In the case of
$p$ polarization, $E_0=E_\parallel^{(1)}$, whereas for $s$
polarization, $E_0=E_\perp^{(1)}$. When the incident field is taken to
be a plane wave, the amplitudes $E_\parallel^{(1)}$ and
$E_\perp^{(1)}$ are constants. It is to be kept in mind that, in order
to lift the sphere from the plate, the positive absorptive force has
to overcome not only gravity, but also the negative surface force
$F_x^{\mathrm{surf}}$ that was calculated in Ref.\ \cite{almaas95} for
the same values of $\{n_1, n_2, n_3\}$ as we consider here. The cases
treated in the two papers are thus directly comparable.

Similarly to Ref.\ \cite{almaas95}, all curves in Figs.\ \ref{fig:3}
-- \ref{fig:5} refer to the case of contact between sphere and plate,
{\it i.e.} to $h=a$. From Eqs.\ (\ref{15}) and (\ref{16}), it is seen
that $A_{lm}$ and $B_{lm}$ contain $\exp(-\beta h)$ as a common
factor. The same property is carried over to the coefficients $c_{lm}$
and $d_{lm}$, according to Eqs.\ (\ref{a21}) and (\ref{a22}). As the
absorptive force is quadratic in the last-mentioned coefficients, it
follows that $Q_x^{\mathrm{abs}}$ at an arbitrary height $h$ is equal
to $\exp[-2\beta (h-a)]$ times the value that can be read off from
Figs.\ \ref{fig:3} -- \ref{fig:5}. No figures need to be worked out to
show the dependence of the force with height above the plate.

In order to interpret the figures physically, it is helpful to
consider a concrete example. Let us relate our discussion essentially
to the situation considered in Section 3: incident plane wave of power
$P=150$ mW, corresponding to $\epsilon_0 E_0^2 a^2=0.13$ pN when the
radius of the sphere is $a=1$ $\mu$m. If the density of the sphere is
2.4 g/$\rm{cm}^3$ (the same density as for glass), the weight of it is
$mg=0.10$ pN. Moreover, $n_3$=1.50, $n_2$= 1.33 (water), whereas the
size parameter is $\alpha=7.9$ for Nd:YAG laser light. Let us look for
the necessary value of $\sigma$ in order to satisfy the condition for
elevation of the sphere, just above the plate. The condition is
$F_x^{\mathrm{abs}}=mg+|F_x^{\mathrm{surf}}|$. Expressed
non-dimensionally,
\begin{equation}
Q_x^{\mathrm{abs}}=\frac{mg}{\epsilon_0 E_0^2 a^2} + |Q_x^{\mathrm{surf}}|.
    \label{57}
\end{equation}
From Fig.\ \ref{fig:3} we read off, when choosing for definiteness the
case of $p$ polarization, $Q_x^{\mathrm{abs}}/(10^{-4}\sigma)=1.7$,
whereas from Fig.\ 4 in \cite{almaas95} we read off
$Q_x^{\mathrm{surf}} =-0.35$ for the same value of $\alpha$. This
leads to a relatively high value of $\sigma$, about 6600 S/m (S$\equiv
\Omega^{-1}$). In our example, the influence of the weight of the
sphere is relatively large because the incident power is so
moderate. If we increase $P$ to 1 watt, we get $\epsilon_0 E_0^2a^2=
0.87$ pN, resulting in a somewhat lower value, $\sigma \approx 2700$
S/m.

For glass, at optical wavelengths, the effective conductivity $\sigma$
is very small. (The static conductivity is extremely small,
$\sigma_{\rm stat} \sim 10^{-12}$ S/m.) Even for higher wavelengths
such as $\lambda_0=8.0\, \mu$m, the value $\sigma=550$ S/m given in
Eq.\ (6) is insufficient to account for the lifting force. For
water, at optical wavelengths, as we have seen in Eq.\ (5), the
effective conductivity is even smaller.  We can thus immediately
conclude that absorption is {\it not} the physical reason for the
elevation of the sphere in the Kawata-Sugiura experiment. It is likely
that there were surfactants on the spheres in this experiment, and
hence, that a theoretical description of the kind given in Ref.\
\cite{lester99} is most appropriate in this case. For other materials
where the conductivities are higher, for instance for carbon
($\sigma_{\rm stat}= 0.77 \times 10^5$ S/m), the absorption-generated
elevation of microspheres should be quite possible physically.  One
must however be aware of the algebraic restriction made in the present
paper to simplify calculations: our calculation relied on the property
that $\sigma/\epsilon_0 \omega \ll 1$ (cf. Eq.\ (\ref{2})). This
restriction made it possible to take the refractive index $n_3$ to be
real, inside the integrals in Section 4. Since $\epsilon_0 \omega
\approx 10^5$ S/m, we can no longer expect high accuracy from the
formalism when $\sigma \gg 10^4$ S/m, at optical frequencies.

Consider finally the remaining figures: Fig.\ \ref{fig:4} shows that
when the refractive index in the sphere increases from $1.50$ to
$1.60$, the absorptive force increases. This is as we would
expect. Moreover, if the difference between interior and exterior
indices increases, the Riccati-Bessel functions cause the force to
become more oscillatory in character. The changes are clearly shown,
if we compare Fig.\ \ref{fig:3} with Fig.\ \ref{fig:5}. Increasing
differences between the refractive indices effectively mean enhancing
the the geometrical-optics properties of the system. There is also a
dependence of the force upon the two different polarization states of
the beam. No simple physical explanation of this dependence seems to
exist.

\section{Conclusions}

We can summarize as follows:

1.  The strategy of the above calculation has been to identify the
evanescent field with the incident field on the sphere. The theory
makes use of wave optics, and is thus applicable even when the
wavelength is of the same magnitude as the sphere diameter.

2.  The main novel development of the present paper in comparison with
Ref.\ \cite{almaas95} is the calculation in Section 4 of the vertical
absorptive force, assuming the refractive index $\bar{n}_3$ in the
sphere to be complex. The calculation rests upon the approximation
that $\sigma/\epsilon_0 \omega \ll 1$, so that the fields inside the
sphere can be evaluated as if the material were a perfect
dielectric. The surface force is thus calculated from Eq.\ (26)
assuming $\bar{n}_3=n_3$ to be real. The absorptive force is
calculated from Eq.\ (30).

3.  The results of Figs.\ \ref{fig:3} -- \ref{fig:5} show that in order
to account for elevation of the micrometer-sphere, the conductivity
$\sigma$ must be at least as large as about $10^3$ S/m. This applies,
for instance, to media like carbon.

4.  For latex or glass spheres, used in the Kawata-Sugiura experiment
\cite{kawata92}, the absorptive force seems to be too weak to account
for the elevation force. Most probably, the observed elevation force
was due largely to impurities or a film on the surface of the sphere,
making it partly conducting. Cf. also Ref.\ \cite{lester99}. The
absorption is sensitive with respect to impurities, and it is at
present difficult to estimate to which degree the absorption is
important in practice.

5.  As a final general remark: the subject of radiation pressure on
particles in the evanescent field is a many-facetted phenomenon which
is not yet fully understood, and it is a very active subject. In
addition to the references given above, the reader is referred also to
recent papers in which the dipole method is made use of, in order to
calculate the force of an evanescent wave over dielectric and metallic
surfaces \cite{chaumet00a,chaumet00b}.

\section{Acknowledgements}

E.A. was supported by NSF through grants DMR97-31511 and
DMR01-04987. Computational support was provided by the Ohio
Supercomputer Center and the Norwegian University of Science and
Technology (NTNU).

\appendix

\section{Expansions of the electromagnetic fields}

\renewcommand{\theequation}{\mbox{\Alph{section}.\arabic{equation}}}
\setcounter{equation}{0}

For convenience, we summarize the expansions for the incident, $(i)$,
the scattered, $(s)$, and the internal, $(w)$, field components. As
mentioned in the main text, subscript 1 refers to the non-absorbing
substrate, subscript 2 to the non-absorbing medium around the sphere
(refractive index $n_2$), and subscript 3 to the absorbing sphere
whose complex refractive index is $\bar{n}_3$. The relative refractive
index is $\bar{n}_{32}=\bar{n}_3/n_2$. Furthermore, $E_0$ and $H_0$
refer to the incoming plane wave in the substrate, the non-dimensional
wave number and distance is $\alpha=k_2 a=n_2\omega a/c$, and
$\tilde{r}=r/a$ respectively.

The formulas below are general. As mentioned in the main text, in the
practical calculations of the coefficients $A_{lm}$ and $B_{lm}$ we
took however the refractive index $\bar{n}_3$ to be real.

\begin{widetext}
\bigskip
\noindent
{\it Incident field}
\begin{eqnarray}
E_r^{(i)} &=& \frac{E_0}{\tilde{r}^2} \sum_{l=1}^\infty \sum_{m=-l}^l
          l(l + 1) A_{lm} \psi_l (\alpha \tilde{r}) Y_{lm}, \label{a1} \\
E_\theta^{(i)} &=& \frac{\alpha E_0}{ \tilde{r}} \sum_{l=1}^\infty
          \sum_{m=-l}^l \left[ A_{lm} \psi_l' (\alpha \tilde{r})
          \frac{\partial Y_{lm}}{\partial \theta} -\frac{m}{n_2}
          B_{lm} \psi_l (\alpha \tilde{r}) \frac{Y_{lm}}{\sin \theta}
          \right], \label{a2} \\
E_\varphi^{(i)} &=& \frac{\alpha E_0}{\tilde{r}} \sum_{l=1}^\infty
          \sum_{m=-l}^l \left[ i m A_{lm} \psi_l' (\alpha \tilde{r})
          \frac{Y_{lm}}{\sin \theta} - \frac{i}{n_2} B_{lm} \psi_l
          (\alpha \tilde{r}) \frac{ \partial Y_{lm}}{ \partial \theta}
          \right], \label{a3} \\
H_r^{(i)} &=& \frac{H_0}{\tilde{r}^2} \sum_{l=1}^\infty \sum_{m=-l}^l
          l(l + 1) B_{lm} \psi_l (\alpha \tilde{r}) Y_{lm}, \label{a4} \\
H_\theta^{(i)} &=& \frac{\alpha H_0}{\tilde{r}} \sum_{l=1}^\infty
          \sum_{m=-l}^l \left[ B_{lm} \psi_l' (\alpha \tilde{r})
          \frac{\partial Y_{lm}}{\partial \theta} + m n_2 A_{lm}
          \psi_l (\alpha \tilde{r}) \frac{Y_{lm}}{\sin \theta}
          \right], \label{a5} \\
H_\varphi^{(i)} &=& \frac{\alpha H_0}{\tilde{r}} \sum_{l=1}^\infty
          \sum_{m=-l}^l \left[ i m B_{lm} \psi_l' (\alpha \tilde{r})
          \frac{Y_{lm}}{\sin \theta} + i n_2 A_{lm} \psi_l (\alpha
          \tilde{r}) \frac{\partial Y_{lm}}{\partial \theta} \right]. 
	  ~~~~~~\label{a6}
\end{eqnarray}

\bigskip
\noindent
{\it Scattered field}
\begin{eqnarray}
E_r^{(s)} \!\! &=& \!\! \frac{E_0}{\tilde{r}^2} \sum_{l=1}^\infty
          \sum_{m=-l}^ll(l+1) a_{lm} \xi_l^{(1)} (\alpha
          \tilde{r})Y_{lm}, \label{a7} \\
E_\theta^{(s)}\!\!  &=&\!\! \frac{\alpha E_0}{\tilde{r}}\sum_{l=1}^\infty
          \sum_{m=-l}^l \left[ a_{lm} {\xi_l^{(1)}}' (\alpha \tilde{r})
          \frac{\partial Y_{lm}}{\partial \theta} - \frac{m}{n_2} \,
          b_{lm} \, \xi_l^{(1)} (\alpha \tilde{r}) \frac{Y_{lm}}{\sin
          \theta} \right], \label{a8} 
\end{eqnarray}
\begin{eqnarray}
E_\varphi^{(s)}\!\! &=&\!\! \frac{\alpha E_0}{\tilde{r}}\sum_{l=1}^\infty
          \sum_{m=-l}^l \left[ i m a_{lm} {\xi_l^{(1)}}' (\alpha
          \tilde{r}) \frac{Y_{lm}}{\sin \theta} - \frac{i}{n_2} \,
          b_{lm} \, \xi_l^{(1)} (\alpha \tilde{r}) \frac{\partial
          Y_{lm}}{\partial \theta} \right], \label{a9} \\
H_r^{(s)} \!\!&=&\!\! \frac{H_0}{\tilde{r}^2}\sum_{l=1}^\infty 
          \sum_{m=-l}^l l(l + 1) \, b_{lm} \, \xi_l^{(1)} (\alpha \tilde{r}) 
	  Y_{lm}, \label{a10} \\
H_\theta^{(s)}\!\! &=&\!\! \frac{\alpha H_0}{\tilde{r}} \sum_{l=1}^\infty
          \sum_{m=-l}^l \left[ b_{lm} {\xi_l^{(1)}}' (\alpha \tilde{r})
          \frac{\partial Y_{lm}}{\partial \theta} + m n_2 a_{lm}
          \xi_l^{(1)} (\alpha \tilde{r}) \frac{Y_{lm}}{\sin \theta}
          \right], \label{a11} \\
H_\varphi^{(s)}\!\! &=&\!\! \frac{\alpha H_0}{\tilde{r}}\sum_{l=1}^\infty
          \sum_{m=-l}^l \left[ i m \, b_{lm} \, {\xi_l^{(1)}}' (\alpha
          \tilde{r}) \frac{Y_{lm}}{\sin \theta} + i n_2 \, a_{lm} \,
          \xi_l^{(1)} (\alpha \tilde{r}) \frac{\partial
          Y_{lm}}{\partial \theta} \right]. ~~~~~~~\label{a12}
\end{eqnarray}

\bigskip
\noindent
{\it Internal field}
\begin{eqnarray}
E_r^{(w)}\!\!\!\! &=&\!\!\!\! \frac{E_0}{\tilde{r}^2} \!\sum_{l=1}^\infty \!\sum_{m=-l}^l
          l(l + 1) c_{lm} \psi_l (\bar{n}_{32} \alpha \tilde{r})
          Y_{lm}, \label{a13} \\
E_\theta^{(w)}\!\!\!\! &=&\!\!\!\! \frac{\alpha E_0}{\tilde{r}} \!\sum_{l=1}^\infty
          \!\sum_{m=-l}^l\!\! \left[ \bar{n}_{32} \, c_{lm} \, \psi_l'
          (\bar{n}_{32} \alpha \tilde{r}) \frac{\partial
          Y_{lm}}{\partial \theta} \!-\! \frac{m}{n_2} \, d_{lm} \, \psi_l
          (\bar{n}_{32} \alpha \tilde{r}) \frac{Y_{lm}}{\sin \theta}
          \right], \label{a14} \\
E_\varphi^{(w)}\!\!\!\! &=&\!\!\!\! \frac{\alpha E_0}{\tilde{r}} \!\sum_{l=1}^\infty
          \!\sum_{m=-l}^l\!\! \left[ i m \bar{n}_{32} c_{lm}  \psi_l'
          (\bar{n}_{32} \alpha \tilde{r}) \frac{Y_{lm}}{\sin \theta} \!-\!
          \frac{i}{n_2}  d_{lm}  \psi_l (\bar{n}_{32} \alpha
          \tilde{r}) \frac{\partial Y_{lm}}{\partial \theta} \right], 
	  \label{a15} \\
H_r^{(w)}\!\!\!\! &=&\!\!\!\! \frac{H_0}{\tilde{r}^2} \!\sum_{l=1}^\infty \!\sum_{m=-l}^l
          l(l + 1)  d_{lm}  \psi_l( \bar{n}_{32} \alpha \tilde{r})
          Y_{lm}, \label{a16} \\
H_\theta^{(w)}\!\!\!\! &=&\!\!\!\! \frac{\alpha H_0}{\tilde{r}} \!\sum_{l=1}^\infty
          \!\sum_{m=-l}^l\!\! \left[ \bar{n}_{32}  d_{lm}  \psi_l'
          (\bar{n}_{32} \alpha\tilde{r}) \frac{\partial
          Y_{lm}}{\partial \theta}\! +\! m n_2  \bar{n}_{32}^2  c_{lm}
          \psi_l (\bar{n}_{32} \alpha \tilde{r}) \frac{Y_{lm}}{\sin
          \theta} \right], \label{a17} \\
H_\varphi^{(w)}\!\!\!\! &=&\!\!\!\! \frac{\alpha H_0}{\tilde{r}} \!\sum_{l=1}^\infty
          \!\!\sum_{m=-l}^l\!\! \left[im\bar{n}_{32}  d_{lm}  \psi_l'
          (\bar{n}_{32} \alpha \tilde{r}) \frac{Y_{lm}}{\sin \theta}\! +\!
          i n_2 \bar{n}_{32}^2  c_{lm}  \psi_l (\bar{n}_{32}
          \alpha \tilde{r}) \frac{\partial Y_{lm}}{\partial \theta}
          \right]. ~~~~~~~\label{a18}
\end{eqnarray}
\end{widetext}

Everywhere, primes denote differentiation with respect to the whole
argument.

\bigskip

\noindent
{\it Relations between coefficients}

\begin{eqnarray}
a_{lm}\!\!\! &=&\!\! \frac{\psi_l' (\bar{n}_{32} \alpha) \psi_l (\alpha) -
       \bar{n}_{32} \psi_l (\bar{n}_{32} \alpha) \psi_l' (\alpha)}
       {\bar{n}_{32} \psi_l (\bar{n}_{32} \alpha) {\xi_l^{(1)}}'\!\!
       (\alpha)\! -\! \psi_l' (\bar{n}_{32} \alpha) \xi_l^{(1)}\! ( \alpha
       )} A_{lm}, \label{a19} \\
b_{lm}\!\!\! &=&\!\! \frac{\bar{n}_{32} \psi_l' (\bar{n}_{32} \alpha) \psi_l
       (\alpha) - \psi_l (\bar{n}_{32} \alpha) \psi_l' (\alpha)}
       {\psi_l (\bar{n}_{32} \alpha) {\xi_l^{(1)}}'\!\! (\alpha)\! -\!
       \bar{n}_{32} \psi_l' (\bar{n}_{32} \alpha) \xi_l^{(1)}\! ( \alpha
       )} B_{lm}, ~~~~~~\label{a20}\\
c_{lm}\!\!\! &=&\!\! \frac{i}{\bar{n}_{32}^2 \psi_l (\bar{n}_{32} \alpha)
       {\xi_l^{(1)}}'\!\! (\alpha)\! -\! \bar{n}_{32} \psi_l' (\bar{n}_{32}
       \alpha) \xi_l^{(1)}\! ( \alpha )} A_{lm}, ~~~~~~~~~\label{a21}
\end{eqnarray}
\newpage

\vspace*{-0.8cm}
\begin{eqnarray}
d_{lm}\!\!\! &=&\!\! \frac{i}{\psi_l (\bar{n}_{32} \alpha) {\xi_l^{(1)}}'\!\!
       (\alpha)\! -\! \bar{n}_{32} \psi_l' (\bar{n}_{32} \alpha)
       \xi_l^{(1)}\! ( \alpha )} B_{lm}. ~~~~~~~~\label{a22}
\end{eqnarray}
Here the Wronskian is $W\{\psi_l, \xi_l^{(1)}\}=i.$

\section{Table of $A_{lm}$ and $B_{lm}$}

We include a table of the numerical values for the real and imaginary
parts of $A_{lm}$ and $B_{lm}$ for an evanescent field. The
coefficients are calculated for $p$ polarization, and we have set
$a=1$ and $e^{-\beta h} = 1$ in Eqs.\ (\ref{15}) and (\ref{16}). For
space considerations, we only give the coefficients up to $l=7$.

\newpage

\begin{tabular}{r r|r r|r r}
 $l$ & $m$ & $\Re\{A_{lm}\}$ & $\Im\{A_{lm}\}$ &  $\Re\{B_{lm}\}$ & $\Im\{B_{lm}\}$ \\ \hline
  1 &  -1 &  7.8951e-14 & -3.5261e-14 &  4.5863e-14 &  1.0269e-13 \\
  1 &   0 & -1.0417e-14 & -2.3324e-14 &  0.0 &  0.0 \\
  1 &   1 & -7.8951e-14 &  3.5261e-14 &  4.5863e-14 &  1.0269e-13 \\
  2 &  -2 & -1.2570e-14 &  5.6142e-15 & -7.3021e-15 & -1.6350e-14 \\
  2 &  -1 &  2.8048e-14 &  6.2800e-14 & -7.8266e-14 &  3.4955e-14 \\
  2 &   0 &  3.0791e-14 & -1.3752e-14 &  0.0 &  0.0 \\
  2 &   1 & -2.8048e-14 & -6.2800e-14 & -7.8266e-14 &  3.4955e-14 \\
  2 &   2 & -1.2570e-14 &  5.6142e-15 &  7.3021e-15 &  1.6350e-14 \\
  3 &  -3 &  2.1752e-15 & -9.7150e-16 &  1.2636e-15 &  2.8292e-15 \\
  3 &  -2 & -7.7601e-15 & -1.7375e-14 &  2.2116e-14 & -9.8777e-15 \\
  3 &  -1 & -5.7659e-14 &  2.5752e-14 & -3.0232e-14 & -6.7690e-14 \\
  3 &   0 &  1.6820e-14 &  3.7661e-14 &  0.0 &  0.0 \\
  3 &   1 &  5.7659e-14 & -2.5752e-14 & -3.0232e-14 & -6.7690e-14 \\
  3 &   2 & -7.7601e-15 & -1.7375e-14 & -2.2116e-14 &  9.8777e-15 \\
  3 &   3 & -2.1752e-15 &  9.7150e-16 &  1.2636e-15 &  2.8292e-15 \\
  4 &  -4 & -3.9427e-16 &  1.7609e-16 & -2.2903e-16 & -5.1281e-16 \\
  4 &  -3 &  1.8142e-15 &  4.0620e-15 & -5.2075e-15 &  2.3258e-15 \\
  4 &  -2 &  2.1681e-14 & -9.6834e-15 &  1.1989e-14 &  2.6843e-14 \\
  4 &  -1 & -2.5411e-14 & -5.6895e-14 &  6.2106e-14 & -2.7738e-14 \\
  4 &   0 & -4.4609e-14 &  1.9923e-14 &  0.0 &  0.0 \\
  4 &   1 &  2.5411e-14 &  5.6895e-14 &  6.2106e-14 & -2.7738e-14 \\
  4 &   2 &  2.1681e-14 & -9.6834e-15 & -1.1989e-14 & -2.6843e-14 \\
  4 &   3 & -1.8142e-15 & -4.0620e-15 & -5.2075e-15 &  2.3258e-15 \\
  4 &   4 & -3.9427e-16 &  1.7609e-16 &  2.2903e-16 &  5.1281e-16 \\
  5 &  -5 &  7.3610e-17 & -3.2876e-17 &  4.2760e-17 &  9.5741e-17 \\
  5 &  -4 & -4.0248e-16 & -9.0116e-16 &  1.1595e-15 & -5.1784e-16 \\
  5 &  -3 & -6.2653e-15 &  2.7982e-15 & -3.5248e-15 & -7.8921e-15 \\
  5 &  -2 &  1.1689e-14 &  2.6172e-14 & -3.1349e-14 &  1.4001e-14 \\
  5 &  -1 &  5.8803e-14 & -2.6263e-14 &  2.6420e-14 &  5.9156e-14 \\
  5 &   0 & -2.3242e-14 & -5.2039e-14 &  0.0 &  0.0 \\
  5 &   1 & -5.8803e-14 &  2.6263e-14 &  2.6420e-14 &  5.9156e-14
\end{tabular}

\begin{tabular}{r r|r r|r r}
 $l$ & $m$ & $\Re\{A_{lm}\}$ & $\Im\{A_{lm}\}$ &  $\Re\{B_{lm}\}$ & $\Im\{B_{lm}\}$ \\ \hline
  5 &   2 &  1.1689e-14 &  2.6172e-14 &  3.1349e-14 & -1.4001e-14 \\
  5 &   3 &  6.2653e-15 & -2.7982e-15 & -3.5248e-15 & -7.8921e-15 \\
  5 &   4 & -4.0248e-16 & -9.0116e-16 & -1.1595e-15 &  5.1784e-16 \\
  5 &   5 & -7.3610e-17 &  3.2876e-17 &  4.2760e-17 &  9.5741e-17 \\
  6 &  -6 & -1.4028e-17 &  6.2652e-18 & -8.1489e-18 & -1.8246e-17 \\
  6 &  -5 &  8.7334e-17 &  1.9554e-16 & -2.5213e-16 &  1.1261e-16 \\
  6 &  -4 &  1.6276e-15 & -7.2694e-16 &  9.2343e-16 &  2.0676e-15 \\
  6 &  -3 & -3.9706e-15 & -8.8903e-15 &  1.0972e-14 & -4.9002e-15 \\
  6 &  -2 & -3.1171e-14 &  1.3922e-14 & -1.6060e-14 & -3.5958e-14 \\
  6 &  -1 &  2.8039e-14 &  6.2781e-14 & -5.7903e-14 &  2.5861e-14 \\
  6 &   0 &  6.0270e-14 & -2.6918e-14 &  0.0 &  0.0 \\
  6 &   1 & -2.8039e-14 & -6.2781e-14 & -5.7903e-14 &  2.5861e-14 \\
  6 &   2 & -3.1171e-14 &  1.3922e-14 &  1.6060e-14 &  3.5958e-14 \\
  6 &   3 &  3.9706e-15 &  8.8903e-15 &  1.0972e-14 & -4.9002e-15 \\
  6 &   4 &  1.6276e-15 & -7.2694e-16 & -9.2343e-16 & -2.0676e-15 \\
  6 &   5 & -8.7334e-17 & -1.9554e-16 & -2.5213e-16 &  1.1261e-16 \\
  6 &   6 & -1.4028e-17 &  6.2652e-18 &  8.1489e-18 &  1.8246e-17 \\
  7 &  -7 &  2.7140e-18 & -1.2121e-18 &  1.5766e-18 &  3.5300e-18 \\
  7 &  -6 & -1.8745e-17 & -4.1970e-17 &  5.4194e-17 & -2.4204e-17 \\
  7 &  -5 & -3.9999e-16 &  1.7865e-16 & -2.2806e-16 & -5.1063e-16 \\
  7 &  -4 &  1.1736e-15 &  2.6278e-15 & -3.2899e-15 &  1.4693e-15 \\
  7 &  -3 &  1.2042e-14 & -5.3781e-15 &  6.4860e-15 &  1.4522e-14 \\
  7 &  -2 & -1.6489e-14 & -3.6919e-14 &  4.0869e-14 & -1.8253e-14 \\
  7 &  -1 & -6.8637e-14 &  3.0655e-14 & -2.5861e-14 & -5.7905e-14 \\
  7 &   0 &  3.1083e-14 &  6.9595e-14 &  0.0 &  0.0 \\
  7 &   1 &  6.8637e-14 & -3.0655e-14 & -2.5861e-14 & -5.7905e-14 \\
  7 &   2 & -1.6489e-14 & -3.6919e-14 & -4.0869e-14 &  1.8253e-14 \\
  7 &   3 & -1.2042e-14 &  5.3781e-15 &  6.4860e-15 &  1.4522e-14 \\
  7 &   4 &  1.1736e-15 &  2.6278e-15 &  3.2899e-15 & -1.4693e-15 \\
  7 &   5 &  3.9999e-16 & -1.7865e-16 & -2.2806e-16 & -5.1063e-16 \\
  7 &   6 & -1.8745e-17 & -4.1970e-17 & -5.4194e-17 &  2.4204e-17 \\
  7 &   7 & -2.7140e-18 &  1.2121e-18 &  1.5766e-18 &  3.5300e-18
\end{tabular}

\end{document}